\documentclass[conference,letterpaper]{IEEEtran}
\usepackage[utf8]{inputenc}
\usepackage{graphicx}
\usepackage[caption=false]{subfig}
\usepackage{array}
\usepackage{amsmath,amsfonts,amssymb}
\usepackage{amsthm}
\usepackage{booktabs}
\usepackage{algorithm}
\usepackage{algorithmic}
\usepackage{multirow, makecell}
\usepackage{flushend}

\newcolumntype{C}[1]{>{\centering\arraybackslash}p{#1}}

\begin{document}
\title{Using Social Media Background to Improve Cold-start Recommendation Deep Models}
\author{\IEEEauthorblockN{Yihong Zhang, Takuya Maekawa, and Takahiro Hara}
\IEEEauthorblockA{Graduate School of Information Science and Technology\\
Osaka University, Osaka, Japan\\
yhzhang7@gmail.com, maekawa@ist.osaka-u.ac.jp, hara@ist.osaka-u.ac.jp}
}
\maketitle

\maketitle
\begin{abstract}
In recommender systems, a cold-start problem occurs when there is no past interaction record associated with the user or item. Typical solutions to the cold-start problem make use of contextual information, such as user demographic attributes or product descriptions. A group of works have shown that social media background can help predicting temporal phenomenons such as product sales and stock price movements. In this work, our goal is to investigate whether social media background can be used as extra contextual information to improve recommendation models. Based on an existing deep neural network model, we proposed a method to represent temporal social media background as embeddings and fuse them as an extra component in the model. We conduct experimental evaluations on a real-world e-commerce dataset and a Twitter dataset. The results show that our method of fusing social media background with the existing model does generally improve recommendation performance. In some cases the recommendation accuracy measured by hit-rate@K doubles after fusing with social media background. Our findings can be beneficial for future recommender system designs that consider complex temporal information representing social interests.
\end{abstract}

\section{Introduction}
In e-commerce, recommender systems can provide personalized recommendation of products and services by discovering hidden user preference from data. Traditionally, the hidden user preference is discovered by techniques such as collaborative filtering and matrix factorization, which are based on users' past interaction with items (purchases or ratings) \cite{sarwar2001item,zhou2012learning,he2017neural}. A problem that attracts attention in recommender system researches in recent years is called cold-start recommendation \cite{lam2008addressing}. In this problem, there is no past record of user-item interaction, because, for example, the user or item is a new one in the system. To make recommendation in such cases, using information other than user-item interaction is necessary. A group of such information is called contextual information \cite{haruna2017context}. Contextual information that has been shown useful in cold-start recommendation include user demographic data \cite{lika2014facing}, item attributes \cite{zhu2019addressing} and item review texts \cite{mcauley2013hidden}.

In this paper, we investigate to what extend social media background can be used as contextual information to improve recommendation accuracy in a cold-start setting. When we mention social media background, we mean the social media text messages that have no explicit connection to either e-commerce users or items. Thus our work is different from previous studies of recommender systems that rely on explicit user connection across e-commerce and social media platforms \cite{zhao2015connecting,alahmadi2015twitter}. Rather, the social media in our study is considered as a temporal background that reflects the general social interests of the moment. We can consider an example scenario: when some local natural disaster happened and received attention in social media, people's interests for disaster prevention would temporarily increase, and disaster prevention products in an e-commerce site would became temporarily fast-selling. This is an actual case observed by our e-commerce partner. Here, what was discussed in social media had an impact in e-commerce user preferences even though users were not linked across platforms. 

Some previous works have shown word usage in social media can be predictive for product sales, even though the effect is relatively weak \cite{zhang2020discovering}. In this work, the social media background is not used as the sole predictive source, but as an addition to an existing recommendation model. We aim to find out how much the recommendation model can be improved by fusing with social media background information. The two technical problems we need to answer are how to represent social media background, and how to fuse them with an existing model. Previously, when social media was used as a background to predict product sales, only crudely-aggregated representation, such as binary sentiment scores, were considered \cite{pai2018predicting,predicting-movie-revenues-twitter}. In this work we provide a finer and more comprehensive representation by using word embeddings, even though they are harder to be incorporated in prediction models.

We base our study on a previous work that proposed a cold start recommender system with deep neural network models \cite{wang2020dnn}. We choose this work because, as we will explain in the Dataset section, the dataset structure in their work is very similar to our work, thus we can adopt their system easily. More specifically, both our work and their work take pre-processing steps to generalize users and items into embeddings based on user and item contextual information. The model in their system is a deep neural network that takes user and item embeddings as inputs. Our task, then, is to add social media information as an extra input to this network.

Our work focus on a specific e-commerce dataset provided by our industry partner. However, we argue that our proposed method has generality that can be adopted and applied in other recommendation application. First, social media platforms such as Twitter make data access available to general public, and data can be easily collected. Second, nowadays more and more recommender systems use deep neural network as the model \cite{zhang2019deep}, and our method to fuse social media information with a specific model can be easily applied to other models. Our main contribution of this work is thus that we propose a novel method to represent and fuse social media background with deep neural network recommendation model. Our experimental results also show that, by fusing with social media background, recommendation accuracy can be steadily improved. To the best of our knowledge, this work is the first to study fusing social media background in cold-start recommendations without explicit links of user accounts.

\section{Related Work}
A number of researches have addressed the problem of cold-start recommendation using contextual information \cite{haruna2017context}. We are most interested in the temporal context, which is close to our work. Cebrian et al. proposed a music recommendation system that used time in the day (morning, afternoon, evening) as the temporal context \cite{cebrian2010music}. Similarly Dias and Fonseca proposed a music recommendation system that considered time in the day, weekday, day of month, etc.,  as well as session information \cite{dias2013improving}. They also clustered songs into latent topics by treating sessions as documents to further improve recommendation. Xiao et al. proposed a probabilistic matrix factorization technique that considers day of week as the temporal context \cite{xiao2018time}. In these works, though, the temporal context is only discrete values of time of the day, and there is no other information associated with such times.

Some works attempted to use social media as the context in recommendation. For example, Alahmadi and Zeng proposed using linked Twitter account to address cold-start recommendation \cite{alahmadi2015twitter}. In order to connect social media to purchase behavior, they explicitly asked e-commerce users to provide their Twitter accounts. Gao et al. studied the problem of location recommendation with location-based social networks \cite{gao2013exploring}. They modeled temporal check-in preferences from users' past check-in records. In their case, both the contextual information and the recommendation target were in the same platform, thus explicit user links were available. In this paper, however, we consider social media purely as a background. We assume no explicit link available for social media and the e-commerce platform. This makes our problem harder, but also increases the generality of our solution.

Outside of recommendation area, social media as a background has been used in different kind of data analysis and applications. For example, Wei el al. have found that Twitter volume spikes could be used to predict stock options pricing \cite{wei2016twitter}. They used the tweets that contained the stock symbols. Asur and Huberman studied if social media chatter can be used to predict movie sales \cite{predicting-movie-revenues-twitter}. They conducted sentiment analysis on tweets containing movie names, and found some positive correlation. Pai and Liu proposed to use tweets and stock market values to predict vehicle sales \cite{pai2018predicting}. They found that by adding the sentiment score calculated from the tweets, prediction model performance substantially increased. Broniatowski et al. made an attempt to track influenza with tweets \cite{broniatowski2015using}. They combined Google Flue Trend with tweets to track municipal-level influenza. Tweets were put through three classifiers to isolate health-related, influenza-related, and case-reporting tweets, and finally the count of relevant tweets was added to the prediction model. These works, however, only used high-level features of social media, such as message counts or aggregated sentiment scores. In this work, we advance the method by representing temporal social media as embeddings, which contain richer information regarding the real-world activities and discussions.

\section{Dataset}
Our study is based on two datasets, including one e-commerce dataset and one social media dataset. In order to show the generality of our method, we will present first our dataset in this section.

\subsection{E-commerce Dataset}
We are provided an e-commerce dataset by our industry partner for the purpose of testing recommendation methods. The dataset contains two types of data, including user browsing data and user purchase data. User browsing data contains the user ID, the URL, and the timestamp for each website visits. It is recorded by a online marketing company. User purchase data records user ID, product id, and the timestamp for each purchase in an e-commerce website. It is recorded by the e-commerce website. Users in browsing data are linked to the users in purchase data. The dataset is of a period of four months, between June and September in 2017. To avoid noises, we select a subset of this dataset that are associated with users who made at least five purchases in the dataset period. There are 6,883 such users, and they are associated with 18,474,185 browsing records and 85,963 purchases.

The products in the e-commerce website, called $deals$, are discount coupons that are made available for a limited period of time, usually between 7 and 14 days. Customer who bought these deals can exchange them with real products and services. The products include several categories of items, including food, cosmetics, home appliances, hobby classes, and travel packages. The number of available deals each day is between 400 to 1000. Each deal in the dataset is associated with a textual description written mostly in Japanese. Since the available periods of the products are short, the market is rapidly changing, and thus the majority of the products can be considered as cold items that have no purchase records in an earlier period.

\subsection{Social Media Dataset}
We obtain a social media dataset by collecting Japanese tweets through Twitter API\footnote{https://developer.twitter.com/en/docs}. To align with the period of e-commerce dataset, we develop a procedure to search past tweets. In addition to time requirement, it is also desirable that the tweets are talking about Japanese domestic affairs, which reflects the background in which the e-commence business was operated. Our procedure is thus as the following. First, we collect a list of Japanese politician Twitter accounts\footnote{Such a list can be found online as political social media accounts are usually public. An example list is provided by the website Meyou with the url https://meyou.jp/group/category/politician/}. From them we remove a few top politician accounts such as Abe Shinzo as they would attract foreign followers. Next we collect the follower of these politicians, who are expected to be Japanese citizens. Then we select from these citizen accounts whose earliest tweets are dated earlier than June, 2017. This is to ensure that the accounts are active during the entire period of the e-commerce dataset. Finally, we collect tweets in the said period from these selected accounts. These tweets become our social media background in this study. In total this dataset contains about 2,464,645 tweets from 33,443 accounts. Intuitively, this social media dataset would only weakly related to user purchase behaviors, since consumer products are not its topic of interests. But messages in this dataset are more similar to typical social media discussions. If we can use this dataset to improve recommendation performance, we can say the totality of social media indeed contains predictive hints.

\section{A Preliminary Study of Social Media Predictive Power}
Previous works have shown that, if the word can be linked to a particular product, word usage in social media can be considered as predictive signals \cite{gruhl2005predictive}. However, when the product is not famous, such as many daily products of minor brands in the e-commerce website in our study, there may not be direct connection between words and products. On the other hand, we have short text descriptions for the products, which may contain key information about the product. For example, all food product using $cheese$ as an ingredient may have the word $cheese$ in the description. We can thus compare common words in product descriptions and social media.

\begin{figure}[ht]
\centering
\subfloat[Steak]{\includegraphics[width=0.18\textwidth]{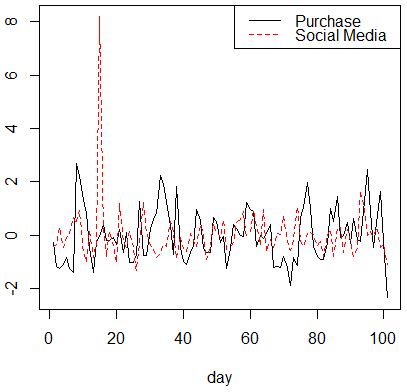}} \quad \quad
\subfloat[Cheese]{\includegraphics[width=0.18\textwidth]{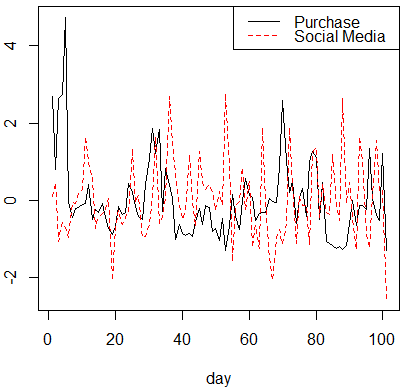}}
\caption{Time series comparison of example words (aggregated by day).}
\label{fig:word}
\end{figure}

Here we provide a preliminary study of two words that are common in social media and product description in order to show the predictive power of social media. We choose two words $steak$ and $cheese$ because they are common words and there is little ambiguity regarding them. We start by counting the frequency of these two words in social media and product purchase. The frequency in social media is counted as the number of tweets containing the word, and in product purchase it is counted as the number of sales of the product containing the word. Since there is a strong seasonal effect that apply to both social media and e-commerce, we remove the seasonal effect by dividing the word count by total word counts:
\begin{equation}
v_w = \frac{count(w)}{\sum_{d\in D} count(d)}
\end{equation}
where $D$ is the entire word vocabulary. After constructing adjusted frequency count for each hour, we obtain time series for social media and product sales shown in Fig \ref{fig:word}.

Pearson's R between two time series are -0.015 for steak, and 0.005 for cheese, which means there is no correlation. However, since we wish to check for predictive power, we perform correlation analysis after shifting social media time series some hours earlier, creating a lag between two time series. The correlations between the two time series give different lags are shown in Table \ref{tab:word}.
\begin{table}[ht]
    \centering
    \caption{Correlation test for lagged time series}
    \label{tab:word}
    \begin{tabular}{c|c c c c c c}
    \hline
         lag&6&5&4&3&2&1  \\ \hline
\multicolumn{7}{l}{steak}\\ \hline
Pearson's R&0.020&0.019&0.040&0.060&0.051&0.011\\
$p$-value&0.324&0.348&0.051&0.003&0.012&0.585\\ \hline
\multicolumn{7}{l}{cheese}\\ \hline
Pearson's R&-0.036&-0.011&-0.014&-0.007&-0.029&0.003\\
$p$-value&0.074&0.587&0.495&0.737&0.152&0.873\\ \hline
    \end{tabular}
\end{table}
As we can see from this table, there is one significant positive correlation between 3-hour-early social media time series and purchase time series for steak. If we set expected significance threshold to $p=0.05$, and perform multiple comparisons correction to get $0.05/6=0.008$, we see that the $p$-value $0.003$ is below the significant threshold. Thus it can be said that earlier usage of the word $steak$ in social media can be used to predict later purchases of products containing the word $steak$ in the e-commerce site. However, for $cheese$, there is no such relationship.

From this preliminary study, we conclude that some words in social media have predictive power for product purchases in e-commerce, while other have not. In this study, we do not pursue the direction of word-level analysis that tries to select predictive words like some previous works \cite{zhang2020discovering}. Instead, we aim to generalize social media information by semantics and let learning models to discover relevant semantic aspects automatically. Our approach has two advantage over word-level analysis. First, as most words used in social media has no clear meaning for an e-commerce product. connection between social media and e-commerce through concrete words is weak. We showed two example words that have little ambiguity, but for most words, the meaning is not clear across social media and e-commerce. Generalize words into semantics by embedding techniques allows a stronger connection to be established. Second, word level representations such as bag-of-words usually requires a large number of dimensions, which are difficult and expensive to incorporate them into existing models. By representing social media with embeddings, we can greatly reduce the dimensionality and make it easier to fuse them with existing models. In the following two sections, we will first present a base recommendation model, and then our approach to fuse social media background with the model.

\section{Base Model}
We select the recommender system proposed by Wang et al. \cite{wang2020dnn} because we our e-commerce dataset structure is very similar to the dataset in their study. Their system can be divided into two main components, namely, a cold-start recommendation model, and user and item embedding learning. In this section, we provide a brief overview of their system, which can be easily apply to our e-commerce data. Please note that they do not use social media data in anyway.

\subsection{Cold start recommendation model}
\begin{figure}[ht]
    \centering
    \includegraphics[width=0.28\textwidth]{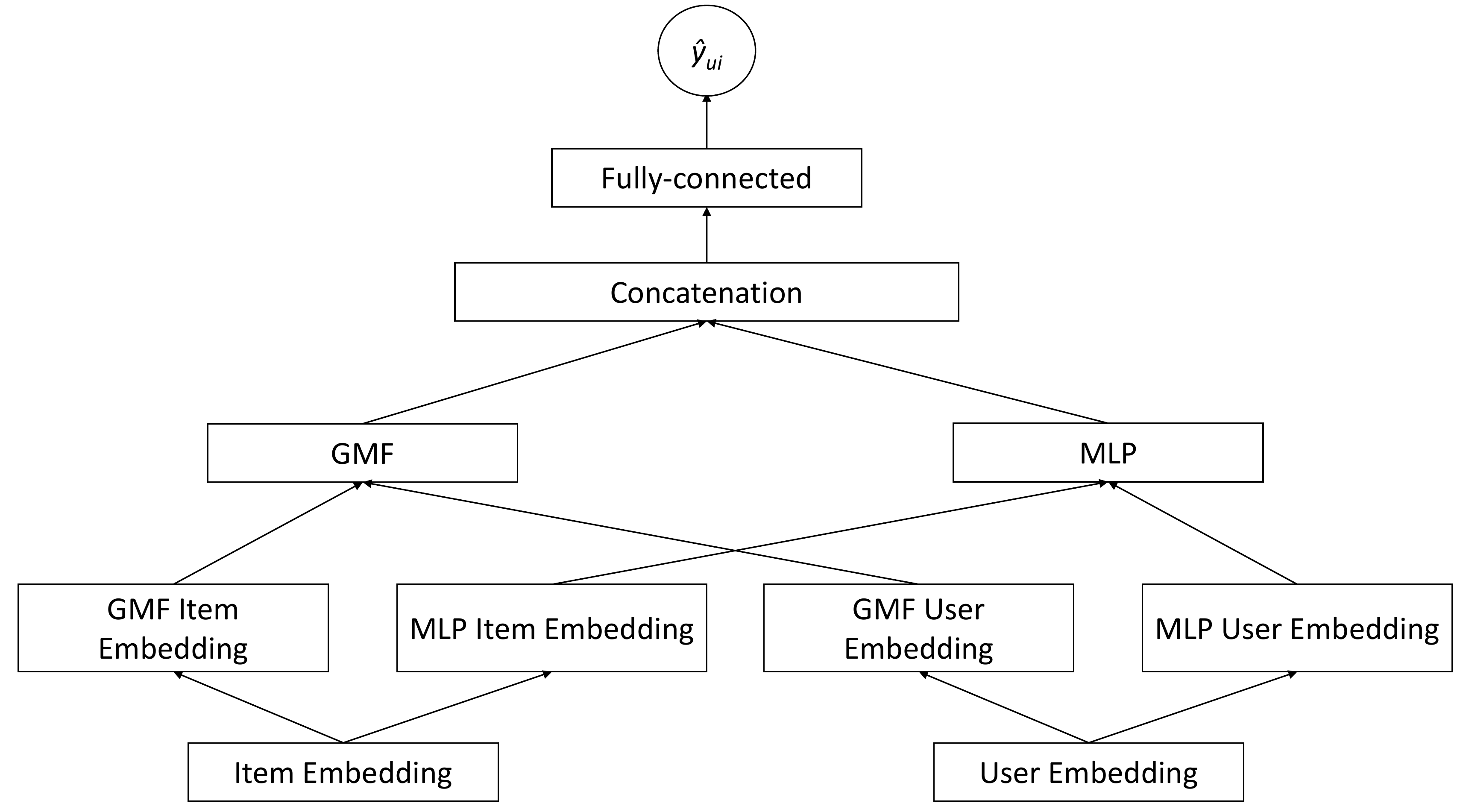}
    \caption{Base recommendation model}
    \label{fig:base}
\end{figure}

Assuming for each item, there is no user-item interaction past records available, and also assuming from the contextual data, embeddings have been learned for users and items. The task of the cold start recommendation model is thus to learn preference relationships between users and items based on their embeddings. Wang et al. generalize a neural matrix factorization (NeuMF) model \cite{he2017neural} that exclude the part of learning embeddings with latent vectors. Their model is shown in Fig. \ref{fig:base}.

This model is an ensemble of generalized matrix factorization (GMF) and multi-layer perceptron (MLP). Two copies of user embeddings and item embeddings are input into the GMF and MLP components, both of which produce an output embedding in their last layer. NeuMF concatenates the two output embeddings and runs them through a fully-connected layer to produce an prediction. The functions in this process are defined as the following:
\begin{equation}
\mathbf{z}_{GMF} = \mathbf{p}_u^G \odot \mathbf{q}_i^G,
\end{equation}
\begin{equation}
\mathbf{z}_{MLP} = a_L(W_L^T(a_{L-1}(...a_2(W_2^T
\begin{bmatrix}
\mathbf{p}_u^M\\
\mathbf{q}_i^M
\end{bmatrix}
+b_2)...))+b_L),
\end{equation}
\begin{equation}
\hat{y}_{ui} = \sigma(\mathbf{h}^T\cdot
\begin{bmatrix}
\mathbf{z}_{GMF}\\
\mathbf{z}_{MLP}
\end{bmatrix}
),
\end{equation}
where $\mathbf{p}_u^G$ and $\mathbf{p}_u^M$ denote user embeddings for GMF and MLP, while $\mathbf{q}_i^G$ and $\mathbf{q}_i^M$ denote item embeddings for the two components. $\hat{y}_{ui}$ denotes prediction results.

Since the dataset contains only observed interactions, i.e., user purchase records of items, when training the model, it is necessary to bring up some negative samples, for example, by randomly choosing some user-item pairs that have no interaction. They defined the loss function as the following:
\begin{equation}
\label{eqn:loss}
L = \sum_{(u, i) \in \mathcal{Y} \cup \mathcal{Y}^-}
y_{ui} \log \hat{y}_{ui} + (1-y_{ui})\log (1-\hat{y}_{ui}),
\end{equation}
where $y_{ui} = 1$ if user $u$ purchased item $i$, and 0 otherwise. $\mathcal{Y}$ denotes observed interactions and $\mathcal{Y}^-$ denotes negative samples.

\subsection{Learning user and item embeddings}
Wang et al. then needed to provide user and item embeddings as the input to above model. Similar to our case, they had a supplementary dataset of user browsing history, as well as textual product descriptions. They proposed two methods to generate user embedding from user browsing histories. The first is through the content of web pages they browsed. They crawled the web pages using the URLs in the browsing log and extract textual information from the HTML tags, such as title, keywords, and description. They further refined the textual content by extracting only the nouns, verbs, and adjective. Then they applied word2vec \cite{mikolov2013distributed} using the skip-gram with negative sampling method to learn embeddings of these words. Each web page was then represented as the average vector of word embeddings. Finally, the user embedding was taken as the average vector of web page vectors in their browsing history.

The second method they used to generate user embeddings from browsing history is through metapath. They transformed their user browsing data into a heterogeneous information network (HIN) \cite{shi2018heterogeneous} so that embeddings can be learned using techniques such as metapath2vec \cite{dong2017metapath2vec}. This method has an advantage over the content-based method in that it captures some semantic relationships, e.g., co-view relationships between users. Specifically, they built a HIN $G = (V, E)$. A node $v \in V$ and a link $e \in E$ are associated with a node type mapping function $\phi(v):V \to T_V$ and a link type mapping function $\psi(e):E \to T_E$, respectively. Based on their data, they defined two type of nodes, the user node $U$, and the website node $W$. A metapath that goes through $U_1, W_1, U_2$ can be created if user $U_1$ and user $U_2$ both browsed the web page $W_1$. They used a random walk algorithm to sample a number of metapaths, and finally applied metapath2vec to generate user embeddings\footnote{here the website embeddings can be generated at the same time, but they are not useful for their work and our work}.

Last, for item embeddings, since the only contextual information about the item is the textual description, they apply word2vec to generate item embeddings from these texts.

\section{Fusing with Social Media Background}
Since social media is dynamically changing, by fusing with social media information, we essentially make the recommendation system time-sensitive. We thus need to complete three tasks. The first is to modify the system so that it becomes aware of the timing. The second is to represent social media background so that they can be input into neural networks. The third is to actually fuse the vector-form social media background with an existing model. We will present our method in this section.

\subsection{Making the Model Time-Aware}
Although it is possible to make recommendation at any moment when a purchase intention is detected, we follow a more realistic scenario by changing the recommendation three times a day, i.e., in the morning, afternoon, and evening, which correspond to hour 10, 16, and 22 of the day. Since in our e-commerce dataset, each purchase is associated with a time, we can modify the target variable to incorporate timing. Specifically, we change $y_{ui}$ to $y_{uti}$ such that $y_{uti} = 1$ if user $u$ purchased item $i$ in the next time segment following $t$, and 0 otherwise. The length of next time segment following hour 10 and 16 is set to 6 hours, and for hour 22 it is set to 12 hours\footnote{The hour of day is taken as the remainder of $t$/24.}. In our dataset, these three time segments separate purchases records evenly, with 29,799, 28,266 and 27,898 purchases in each of the three segments.

It is an important problem to determine whether the user has purchase intention at hour $t$ or not, before making the recommendation. Here we assume this information is already obtained. We use the training label $y_{uti}$ such that there is guaranteed to be an item $i$ that user $u$ will purchase for time $t$. In other words, time $t$s when user $u$ made no purchase at all are ignored. In a complete recommendation system, though, a component is required to make this decision, and we will consider it in a future work.

\subsection{Representing Social Media Background}
We design a method that aims to capture the $change$ or $tendency$ of the social media background that comes from interesting phenomenon happening in real-world. When something interesting or stimulating happens, some topics in social media may become trending. When this happens, we say certain social media aspects are \emph{emerging}. We capture this emergence by observing the change of word frequency. The frequency of the social media words is taken as the count of messages that contains the word. Thus we first obtain the frequency table of social media words against time units. We then devise a method for emergence detection based on word frequencies. Follow an approach of previous works on social media event detection \cite{chen2013emerging}, our method involves a foreground and a background. Suppose the period for foreground is $fp$, and for background is $bp$, so that word frequencies in these periods are $F_{fp}=\{f_{t-fp},...,f_{t-1}\}$ and $F_{bp}=\{f_{t-fp-bp},...,f_{t-fp-1}\}$. We set $inc_{fp}$ to $True$, if $f_{t-1} > \mu(F_{fp})$, where $\mu(\cdot)$ is the mean function, i.e., the frequency in the last day in the foreground period increases compared to the mean of foreground period, and $False$ otherwise. Similarly we set $inc_{bp}$ for the background period. Finally the emergence $e_t$ of the word at time $t$ is set as:
\begin{equation}
e_t =
\begin{cases}
1, \quad \text{if } inc_{fp} \textit{ OR } (inc_{bp} \textit{ AND } \mu(F_{fp}) > \mu(F_{bp}))\\
0, \quad \text{otherwise}
\end{cases}
\end{equation}

With this formula, we aim to capture two phases of surges of words in social media. First, $inc_{fp}$ captures a new surge. Second, $inc_{bp} \textit{ AND } \mu(F_{fp}) > \mu(F_{bp})$ captures the sustenance of a previous surge. Both phases can be considered as a part of an emergence. With this calculation, we obtain for each time unit the emerging words in product sales and social media.

We then generalize the emerging words into semantic meanings, basically using the word2vec method. Based on an implementation made available online\footnote{https://github.com/philipperemy/japanese-words-to-vectors}, we learn a set of Japanese word embeddings using Wikipedia. Then we match words in social media with these word embeddings. We use a natural language processing package called kuromoji\footnote{https://github.com/atilika/kuromoji} to process the Japanese text. The package can effectively perform segmentation and part-of-speech (POS) tagging for Japanese text. After POS tagging, we select only nouns to represent the information in the text. To represent a time unit which consists of a number of emerging words from social media, we take the average vector of the embeddings of these words. Practically, we take hour as the time unit, and obtain hourly embeddings representing social media background with the above method.

\subsection{Fusing Social Media Background with the Model}
A simple way to fuse social media background with the model is by concatenating the embedding with outputs of a middle layer. Specifically, where as $y_{uti}$ indicates user purchase in the next time segment following time $t$, the social media background considered at time $t$ are the embeddings between $t-k$ and $t-1$. We can obtain social media embedding at time $t$, $S_t$, as the average embedding of hourly embedding for the hours in the previous time segment before $t$. A possible position to add social media background input is in the concatenation layer where the outputs of GMF and MLP are jointed. If we do so, the prediction from the final layer becomes
\begin{equation}
\label{eqn:average}
\hat{y}_{uti} = \sigma(\mathbf{h}^T\cdot
\begin{bmatrix}
\mathbf{z}_{GMF}\\
\mathbf{z}_{MLP}\\
S_t
\end{bmatrix}
),
\end{equation}
and the loss function defined in Equation (\ref{eqn:loss}) is modified so that the temporal aspect is considered
\begin{equation}
L = \sum_{(u, t, i) \in \mathcal{Y} \cup \mathcal{Y}^-}
y_{uti} \log \hat{y}_{uti} + (1-y_{uti})\log (1-\hat{y}_{uti}).
\end{equation}

However, above method takes the same social media embedding for any user-item pair in the same time unit. As we have shown in the preliminary analysis, given one item, only some aspects in the social media are relevant. Therefore, the model can be improved if we can select some aspects in social media based on the item currently under consideration.

In recent years, the attention mechanism in deep learning has been shown to be helpful by allowing the model to focus on some aspects of input data \cite{vaswani2017attention}. The goal of an attention module is to produce a weighted average of candidate embeddings of a reference source, called keys, based on their relationships with a query embedding. In our case the keys are hourly social media embeddings and query are item embeddings. The output of an attention module is thus a context vector $c_i$ for item $i$
\begin{equation}
\mathbf{c}_i = \sum_j a_{ij} s_j
\end{equation}
where $s_j$ is the social media embedding in hour $j$, and $a_{ij}$ is called attention weights. The attention weights can be generally obtained using the following formula
\begin{equation}
\mathbf{a}_i = \text{softmax} f_{att}(h_i, s_j)
\end{equation}
where $h_i$ is the embedding of item $i$, and $f_{att}$ is an attention score function calculated on $h_i$ and $s_j$. Several ways has been proposed to calculate attention weights. In this paper we choose a simple approach called the $general$ attention function \cite{luong2015effective}. Basically, it is calculated as
\begin{equation}
f_{att}(h_i, s_j) = h_i^\intercal W s_j
\end{equation}
where W is a randomized weight matrix. Although theoretically simple, it has been shown that this function can capture relevance of keys with respect to the query.

\begin{figure}[ht]
    \centering
    \includegraphics[width=0.48\textwidth]{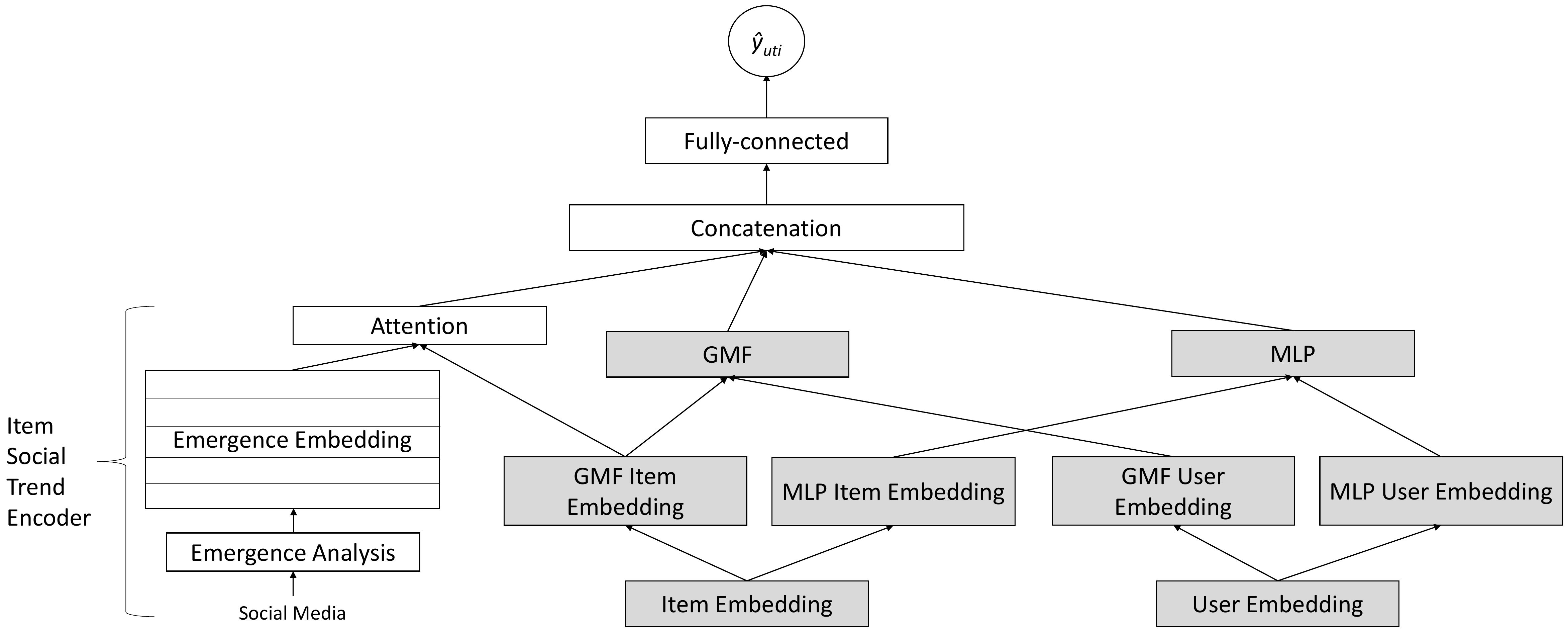}
    \caption{Fusing social media background with an existing model}
    \label{fig:add}
\end{figure}

We consider that the attention mechanism is what we need for finding relevant social media aspects with respect to an item. We thus design the extension so that the social media emergence matrix over $k$ hours is input into an attention component together with one of the item embeddings. In experimental analysis we choose the item embedding output by the GMF component, but the MLP item embedding provides similar results. We call this part of extension Item Social Trend Encoder (ISTE). The model after adding the extension is shown in Fig. \ref{fig:add}. The gray components are unchanged components in the original model, while the bright components are new additions or components that have changed because of the fusion.

With this extension, the concatenation layer takes the output of the encoder and thus becomes
\begin{equation}
\label{eqn:iste}
\hat{y}_{uti} = \sigma(\mathbf{h}^T\cdot
\begin{bmatrix}
\mathbf{z}_{GMF}\\
\mathbf{z}_{MLP}\\
\mathbf{c}_i
\end{bmatrix}
),
\end{equation}

Essentially, fusing with social media background this way allows the final layers of the model to learn the latent relationship between social media background and user-item pairs. More specifically, the co-occurrence of social media aspects and positive/negative instances will be captured. In the case of word $steak$ for example, when the model receives many times the co-occurrence of the emerging word $steak$ and the purchase behavior of products containing the word $steak$, it will reinforce this pattern. By allowing attention on different time in the day, we also capture context that could have different delays in the causality between social media and purchase.

\section{Experimental Evaluation}
We conduct experiments to test the effect of fusing with social media background on product recommendation performance. For this purpose, we implement the original model and the new model after fusing with social media components. Then evaluation on our dataset is conducted on these two models. In this section we will present the experimental setup and discuss the results.

\subsection{Implementation Details}
We generate embeddings of the original model, including content embeddings, metapath embeddings, and item embeddings, using the Gensim library\footnote{https://radimrehurek.com/gensim/}. We set the dimension of these embeddings to 200. When generating metapath, we set random walk length to 160, and initiate 30 walks from each node. For social media embeddings, we set the dimension to 50, considering computational efficiency. 

We use Pytorch\footnote{https://pytorch.org/} to implement the original and the new models. Model parameters are randomly initiated. We follow the original model and use a tower pattern for the MLP component, which halves the layer size for each successive higher layer. The sizes of three layers in the MLP are thus [200, 100, 50]. The output size of GMF is set to 50. The size of fully connected layer is set to 100. The number of hours, $k$, to consider as embeddings in time $t$ is set to 24.

We use the Adam optimizer with an initial learning rate of 0.001. We run 50 training epochs for each model, before which model performances generally become stable. When training the model, we randomly sample 4 negative instances for each positive instance. For the original model, the negative instances (u, i) are user $u$ and item $i$ that have no interaction. While for the new model, the negative instances (u, t, i) are user $u$ and item $i$ that have no interaction for time $t$.

\subsection{Evaluation Settings}
We divide that dataset into a training set and a test set. The training set is of a period from June 1 to September 16, 2017, and test set is between September 17 to 30, 2017, a period of two weeks. We remove what are so called \emph{free items}, that are time-limited discount coupons of 0 price, which anyone can get during the active period without paying a fee. These items occupy a large portion of dataset, but they do not reveal user-item preferences, and their relationship with social media background, so we consider them noises. After removing the free items, the number of users, items, and interactions are shown in Table \ref{tab:dataset}.
\begin{table}[ht]
    \centering
   \caption{Training and test dataset statistics}
    \label{tab:dataset}
    \begin{tabular}{l|C{1.5cm}C{1.5cm}C{2cm}}
    &no. users&no. items&no. interactions\\ \hline
training&2,988&4,146&7,003\\
test&605&583&937\\
    \end{tabular}
\end{table}

The number of available items in a time segment is roughly between 400 and 1,000. Considering this and the consistency of the evaluation, for each interaction in the test dataset (positive instances), we randomly sample 399 negative instances from items available of the same time segment. Combining the positive and negative instances, we have 400 candidate items in each recommendation. 

We use hit-rate (HR) and Normalized Discounted Cumulative Gain (NDCG) 
to measure recommendation performance. HR@K is calculated as
\begin{equation}
HR@K = \frac{\text{number of hits in top K recommendation}}{\text{number of recommendations}}.
\end{equation}
HR@K measures whether the correct item is in the recommended items, but it does not consider the rank of the item. NDCG on the other hand counts the position of correct item. It is calculated as:
\begin{equation}
NDCG@K = \sum_{i=1}^K \frac{2^{r_i} - 1}{\log_2(i+1)},
\end{equation}
where $r_i=1$ if the correct item is ranked in the $i$-th position, and 0 otherwise. NDCG@K will be higher if the correct item is ranked higher in recommended items. To simulate a realistic scenario, where the recommended items are shown in a single web page to e-commerce website visitors, we choose a number of K values between 1 and 10.

Since the purpose of this study is to examine the effect of fusing social media background with an existing system, we will refrain from comparing many other baselines. We provide the result of one new baseline, called \emph{previous popular} (PrevPop), for which items are ranked based on their number of sales in the previous time segment at the time of recommendation. This baseline is only possible because we have modified the model to make it time-aware. Items not available in the previous time segment is treated as having zero sales. We also compare two variation of our extension, one does not use attention and use Equation (\ref{eqn:average}) for the concatenation layer (average). The other use the ISTE and Equation (\ref{eqn:iste}) for the concatenation layer (ISTE).

\subsection{Result Discussion}
\begin{table*}[ht]
    \centering
    \caption{Effect of fusing with social media background on recommendation accuracy}
    \label{tab:acc}
    \begin{tabular}{l| C{1cm}C{1cm}C{1cm}C{1cm}C{1cm} | C{1cm}C{1cm}C{1cm}C{1cm} }
&\multicolumn{5}{c}{HR@K (percentage)}&\multicolumn{4}{c}{NDCG@K}\\
K&1&2&3&5&10&2&3&5&10\\ \hline
PrevPop&0.44&0.55&0.66&1.20&1.97&0.005&0.006&0.008&0.010\\ \hline
content&0.55&1.31&1.75&3.50&7.65&0.010&0.012&0.020&0.033\\
content+SM average&0.98&2.19&2.95&4.26&7.76&0.017&0.021&0.027&0.038\\
\quad absolute increment&+0.44&+0.87&+1.20&+0.77&+0.11&+0.007&+0.009&+0.007&+0.005\\
\quad relative increment&80.0\%&66.7\%&68.7\%&21.9\%&1.4\%&69.3\%&70.3\%&35.7\%&14.6\%\\
content+SM ISTE&1.31&2.19&3.28&4.26&7.98&0.019&0.024&0.028&0.040\\
\quad absolute increment&+0.77&+0.87&+1.53&+0.77&+0.33&+0.83&+1.16&+0.85&+0.71\\
\quad relative increment&140.0\%&66.7\%&87.5\%&21.9\%&4.3\%&81.0\%&93.1\%&43.3\%&21.4\%\\
\hline
metapath&0.55&1.31&1.86&4.26&7.21&0.010&0.013&0.023&0.032\\
metapath+SM average&0.87&2.30&3.39&4.70&7.98&0.018&0.023&0.029&0.039\\
\quad absolute increment&+0.33&+0.98&+1.53&+0.44&+0.77&+0.007&+0.010&+0.006&+0.007\\
\quad relative increment&60.0\%&75.0\%&82.4\%&10.3\%&10.6\%&72.1\%&77.9\%&25.3\%&20.2\%\\
metapath+SM ISTE&1.53&2.62&3.72&4.81&8.09&0.022&0.028&0.032&0.042\\
\quad absolute increment&+0.98&+1.31&+1.86&+0.55&+0.87&+1.19&+1.46&+0.92&+1.01\\
\quad relative increment&180.0\%&100.0\%&100.0\%&12.8\%&12.1\%&115.7\%&112.4\%&40.2\%&31.3\%\\
   \end{tabular}
\end{table*}

Recommendation performances measured as HR@K and NDCG@K of different models are shown in Table \ref{tab:acc}. The first row is the result of PrevPop baseline. The second and third rows show the results of models using content or metapath as the user embedding, without and with the social media component. For each measurement of models with social media, the absolute and relative increments by fusing with social media are shown below that measurement. NDCG@1 results are not shown because they are the same as HR@1.

The main insight from these results is that fusing with social media background using our proposed method generally improves the recommendation prediction. Especially, when K set to small values between 1 and 3, the relative increments by ISTE models are large, between 100\% and 180\% when use with metapath. When K is set to larger values of 5 and 10, the relative increments are small, but nevertheless show steady increase. Since it is shown that it is more effective to fuse with social media when K is small, which means correct items are more likely to be ranked higher, NDCG results are better even when K is large. NDCG@K increased by more than 100\% for the metapath model, when K is less than 3, and still more than 30\% when K is 5 or 10. The reason behind good performance for smaller K value may be that, some products are particular sensitive to social interests, and thus can be picked up and moved to a higher position in the rank by models fused with social media information. But other products do not have such sensitivity, thus these models can only have limited effectiveness. When comparing simple average social embedding method and ISTE, we see that ISTE method with attention has significant advantage. Especially when K=1 and use the metapath, ISTE achieves 120\% more increment than the average method. 

Similar to the finding of the paper proposing the original model, the model with metapath user embeddings is slightly better than the model with content user embedding. All models with and without social media are better than the baseline method PrevPop. The PrevPop is nevertheless better than the theoretic random baseline, for which HR@K is $(1/400)\times K$. Such results show that there is practical value of deploying our model in actual recommender systems.



\section{Conclusion}
In this paper, we investigate the question whether social media background can be used as extra contextual information in cold-start recommendation models. Based on an existing deep neural network model, we propose a method to represent social media background as embeddings and input them into the middle-layer of the network. By doing this, we also make the model time-aware. Experimental evaluations with real-world e-commerce and social media datasets show that this method is successful, with steadily improved performance achieved by models fused with social media.  However, it is still difficult to tell which social media contexts are predictive for which products. This is difficult to explain by examining the model itself. To make our method more suitable for practical deployment, in the future, we will find some ways to make the relationships between social media word usages and product sales more clear. Nevertheless, we consider that the finding in this work is useful for future recommender system designs that consider temporal factors.


\section*{Acknowledgement}
This research is partially supported by JST CREST Grant Number JPMJCR21F2.


\end{document}